# Spin Seebeck effect and thermal spin galvanic effect in $Ni_{80}Fe_{20}$/p-Si bilayers

Ravindra G. Bhardwaj[1], Paul C Lou[1] and Sandeep Kumar[1,2*]

[1] Department of Mechanical Engineering, University of California, Riverside, CA 92521, USA

[2] Materials Science and Engineering Program, University of California, Riverside, CA 92521, USA

Abstract

The development of spintronics and spin-caloritronics devices need efficient generation, detection and manipulation of spin current. The thermal spin current from spin-Seebeck effect has been reported to be more energy efficient than the electrical spin injection methods. But, spin detection has been the one of the bottlenecks since metals with large spin-orbit coupling is an essential requirement. In this work, we report an efficient thermal generation and interfacial detection of spin current. We measured a spin-Seebeck effect in $Ni_{80}Fe_{20}$ (25 nm)/p-Si (50 nm) (polycrystalline) bilayers without heavy metal spin detector. The p-Si, having the centosymmetric crystal structure, has insignificant intrinsic spin-orbit coupling leading to negligible spin-charge conversion. We report a giant inverse spin-Hall effect, essential for detection of spin-Seebeck effect, in the $Ni_{80}Fe_{20}$/p-Si bilayer structure, which originates from Rashba spin orbit coupling due to structure inversion asymmetry at the interface. In addition, the thermal spin pumping in p-Si leads to spin current from p-Si to $Ni_{80}Fe_{20}$ layer due to thermal spin galvanic effect and spin-Hall effect causing spin-orbit torques. The thermal spin-orbit torques leads to collapse of magnetic hysteresis of 25 nm thick $Ni_{80}Fe_{20}$ layer. The thermal spin-orbit torques can be used for efficient magnetic switching for memory applications. These scientific breakthroughs may give impetus to the silicon spintronics and spin-caloritronics devices.

The performance of thermoelectric semiconductors, especially commercially available, has been stagnant for years. The materials that show increase in thermoelectric performance require complex and scarce (rare earth) elements. An innovative approach to improving thermoelectric energy storage and conversion is the spin dependent thermoelectric energy conversion using spin Seebeck effect (SSE), anomalous Nernst effect (ANE) and spin Nernst effect (SNE), which will bring efficiencies because pure spin current, as opposed to charge current, is believed to be dissipationless[1]. The discovery of Spin Seebeck effect (SSE) by Uchida *et. al.* has led to significant progress in ongoing research on generation of pure spin current, a precession of spins or flow of electrons with opposite spins in opposite directions, over a large distance in spintronic devices due to applied temperature gradient in ferromagnetic (FM) materials[2-4]. The SSE can be an efficient way to produce low cost and large memory spintronics devices[5]. The SSE is observed in ferromagnetic metals[3,6-11], semiconductors[12-15], insulators[16-22] and even in half metallic Heusler compounds[23]. In the spin caloritronics studies, homogenous temperature gradient as well as length scale dependent temperature gradient is established to study the interplay of spin degrees of freedom and temperature gradient in the magnetic structures[22]. There are two universal SSE device configuration, longitudinal spin Seebeck effect (LSSE) and transverse spin Seebeck effect (TSSE) in which in-plane external magnetic field and temperature gradient is applied in the plane of the sample to measure the SSE[22]. In LSSE[11], a spin current is generated parallel to the temperature gradient as opposed to the spin current is perpendicular to the temperature gradient in TSSE[4,5,21]. The spin current generated in a FM material is detected by inverse spin-Hall effect (ISHE) in a high spin orbit coupling metals (Pt, W and Ta) in contact with FM [3,5,21]. The ISHE voltage $E_{ISHE}$ generated perpendicularly to the magnetization $M$ is given by equation,

$$E_{ISHE} = (\theta_{SH}\rho)J_S \times \sigma \qquad (1)$$

Where, $\theta_{SH}, \rho, J_S$ and $\sigma$ are spin-Hall angle, electrical resistivity of paramagnetic metal, longitudinal spin current due to SSE, and spin polarization vector parallel to $M$ [3,7]. The thermoelectric energy conversion from spin current depends on efficient spin to charge conversion. Currently, the primary material for spin to charge conversion is Pt due to its large spin Hall angle, which inhibits the further scientific research in spin thermoelectric conversion behavior. The SSE is enhanced due to phonon drag[24] and phonons drive the spin redistribution [13]. The spin-phonon coupling can provide an able platform to engineer spin dependent thermoelectric conversion. To make the spin mediated thermoelectric energy conversion a reality, we need to discover earth abundant material/interfaces for giant SSE/ANE/SNE and efficient spin to charge conversion. In this work, we report the experimental measurement of giant SSE and thermal spin galvanic effect (SGE[25-27]) in the $Ni_{80}Fe_{20}$/p-Si (poly) bilayers. The spin-phonon coupling in p-Si leads to giant enhancement in SSE at the $Ni_{80}Fe_{20}$/p-Si (poly) bilayer and SHE in p-Si leads to giant spin-orbit torque (SOT), which can be used for SOT based memory applications.

We developed an experimental setup to measure the longitudinal SSE. In the experimental setup, we use Pt heater to create the temperature gradient across the $Ni_{80}Fe_{20}$/p-Si (poly) bilayer specimen as shown in the Figure 1-a. This temperature gradient will lead to spin current in the bilayer and will allow us to measure the spin mediated thermoelectric behavior. An AC bias across the Pt heater creates the temperature gradient. We measure the first harmonic and third harmonic response across the heater to quantify the temperature gradient between the heater and the Si substrate. The SSE, ANE and SNE are measured from the second harmonic response across the $Ni_{80}Fe_{20}$/p-Si bilayer specimen. We use the micro/nanofabrication techniques to make the experimental setup as shown in Figure 1 b. To fabricate the experimental setup, we take a Si wafer and deposit 300 nm of silicon oxide using plasma enhanced chemical vapor deposition (PECVD). We, then, deposit the $Ni_{80}Fe_{20}$/p-Si (poly) bilayer (blue color) using the RF sputtering as shown in Figure 1 b. The p-Si target is B-doped having resistivity of 0.005-0.01 Ω-cm. We sputter 50 nm MgO (green color) to electrically isolate the heater and the specimen. We then deposit Ti (10 nm)/Pt (100 nm) (pink color), which acts as a heater.

(Figure 1)

The experimental measurement is carried inside a quantum design physical property measurement system (PPMS). For energy conversion applications, the thermoelectric behavior should be robust at higher temperatures. We applied a 20 mA-5 Hz of heating current across the outer two electrodes of Pt heater starting at 400 K. We then measured the second harmonic response as a function of applied magnetic field in z-direction and y-direction as shown in Figure 1 c. For the magnetic field in y-direction, the field is perpendicular to the direction of temperature gradient and we observe a large second harmonic response, which may be related to the ANE/SSE. But, we observe an equally large signal when the magnetic field is applied along z-direction (field parallel to the temperature gradient). The $Ni_{80}Fe_{20}$ thin films have in-plane magnetic easy axis and out of plane hard axis[28,29], which are verified from the magnetoresistance measurement as shown in Supplementary Figure S1 and anisotropic magnetoresistance (AMR) in Supplementary Figure S2. The second harmonic response in z-direction is attributed to the hard axis magnetization. We, then, measured the second harmonic response as a function of heating power at 400 K as shown in Figure 1 d. We observe linear relationship between the heating power and the second harmonic response[30] as expected.

(Figure 2)

We then measured the second harmonic responses as a function of magnetic field (from 1000 Oe to -1000 Oe) and applied current of 15 mA to 50 mA at 300 K as shown in Figure 2 a-d. We observe linear second harmonic response (comprising of ANE and SSE) as a function of heating power. Surprisingly, we observe that the magnetic hysteresis of second harmonic response for field along z-direction collapses as the heating current is raised to 50 mA. At 30 mA of heating current at 300 K, we estimate the increase in temperature at heater to be ~20.84 K from the third harmonic measurement. This lead us to believe that the observed collapse of hysteresis for z-direction at 300 K cannot arise due to heating effect only since the collapse of hysteresis in z-direction is not observed at 400 K. This behavior indicates existence of additional spin current from p-Si (Poly) to $Ni_{80}Fe_{20}$ layer. This additional spin current leads to spin-orbit torque and resulting change in the hysteresis behavior. For the magnetization perpendicular to the plane of interface, ANE and SSE will not be observed since $M||\Delta T$ and $J_s||\sigma$ respectively.

In order to decouple the contributions of ANE, SSE and to discover the origin of SOT, we measured the second harmonic response for an applied magnetic field (2T) rotated in the xy, zy and zx-plane as shown in Figure 3 a, where temperature gradient is along the z-direction. We observe a sine behavior attributed to the SSE in the xy-rotation. The angular dependence in zx-plane is observed to be cosine and zy-plane shows combined sine and cosine behavior. These measurements led us to believe there is a second thermoelectric effect in the bilayer thin films that is giving rise to the cosine second harmonic response in addition to the out of plane magnetic field dependent behavior reported in Figure 2. This behavior is similar to the SGE response reported in Fe/GaAs structures due to Rashba effect[25]. The thermal SGE, in this study, will lead to charge current across the bilayer specimen causing the second harmonic response. We will like state that the ANE coefficient in $Ni_{80}Fe_{20}$ is extremely small (4.8 nV/K[31]). In addition, the specimen size in our measurement is 160μm X 12 μm, which is relatively very small as compared to sample area for the ANE measurements reported[31,32]. The ordinary Nernst effect (ONE) is not considered in this study because the ONE does not give rise to the switching behavior observed in this work. These measurements lead to two challenges in the interpretation of the results. First, the SSE measurement requires inverse spin-Hall effect (ISHE) to convert the spin current into voltage. While the SHE has been reported in p-Si[33,34] but the spin-Hall angle of p-Si is negligible and may not lead to observable signal. To address the first challenge, we hypothesize that the ISHE occurs due to Rashba spin orbit coupling at the $Ni_{80}Fe_{20}$/p-Si interface.

The second challenge is to uncover the origin of SOT observed in this study. As stated earlier that cosine behavior is attributed to the thermal SGE, which causes the second harmonic response while the magnetization and temperature gradient are parallel to each other. But, this behavior should not lead to SOT observed in Figure 2. We propose a two-step process that will lead to SOT observed in this study. The first step is thermal SGE where tunneling of spin polarized electrons across the interface in z-direction, having polarization in z-direction as well, lead to charge current parallel to the interface in x-direction due to inverse Rashba-Edelstein effect, which can be written as:

$$J_c^x = \lambda_{SGE} J_s^{z,\sigma_z} \quad \text{[25,35]} \tag{2}$$

where $\lambda_{SGE}$ is "effective thickness" of spin orbit layer[25,35]. Since the spin current is a function of temperature gradient, this equation can be written as:

$$J_c^x \propto \lambda_{SGE}(T_{FM} - T_{Si}), \tag{3}$$

where $T_{FM}$ and $T_{Si}$ are temperature of ferromagnetic layer and temperature of Si layer respectively. The Rashba potential leads to spin precession causing a projection of the polarization in y-direction, which then leads to ISHE and charge current in x-direction[25,35]. The field like SOT acts along $\hat{m}\times\hat{\sigma}$ and damping like torque acts along $\hat{m}\times(\hat{m}\times\hat{\sigma})$, where $\hat{m}$ and $\hat{\sigma}$ are the unit vectors of magnetization and spin polarization respectively[36]. For $\hat{m}$ acting in the z-direction, the spin polarization ($\hat{\sigma}$) vector has to be in the plane of the thin film for the SOT. In the second step, the interfacial charge current leads to SHE due to Rashba spin-orbit coupling causing an inverse spin current from p-Si to $Ni_{80}Fe_{20}$ layer having spin polarization in the plane of the thin film. The charge to spin conversion relationship can be written as:

$$J_s^{z,\sigma_y} = \theta_{SHE} J_c^x, \qquad (4)$$

where $\theta_{SHE}$ is spin-Hall angle. The spin current entering the $Ni_{80}Fe_{20}$ layer can be considered as magnetization entering and exiting the ferromagnetic layer, which will cause the spin orbit torque. The spin current causing the SOT can be related to the temperature gradient through the following approximate equation:

$$SOT \propto J_s^{z,\sigma_y} \propto \theta_{SHE} \lambda_{SGE} (T_{FM} - T_{Si}) \qquad (5)$$

The SOT characterization requires application of electric current across the specimen and measurement of first and second harmonic Hall responses. In this study, the $Ni_{80}Fe_{20}$ thin film is two orders of magnitude more conducting than the p-Si (poly) layer. Hence, the SOT observed in this study is not quantifiable with current techniques since it is of thermal origin. But, the SOT leads to collapse of hysteresis in a 25 nm $Ni_{80}Fe_{20}$ thin film as compared to the few nanometer films used in the SOT studies[37-40] and only earth abundant materials are used. While we propose that the second harmonic response for out of plane magnetic field is due to Rashba effect mediated thermal SGE but other mechanisms may also be present, which can lead to better understanding of the observed measurements.

(Figure 3)

Now, we needed to quantify the LSSE at the $Ni_{80}Fe_{20}$/p-Si (poly) interface. The efficiency of converting spin current-voltage at interface of bilayer in a LSSE device is given by [41],

$$S_{LSSE} = \frac{E_{ISHE}}{\nabla T} = \frac{V_{ISHE} t_{FM}}{w_{NM} \Delta T} \tag{6}$$

Where, $V_{ISHE}$ is the electric voltage measured due to ISHE by paramagnetic metal or normal metal (NM), $t_{FM}$ is thickness of FM material, $w_{NM}$ is the distance between electrical contact in NM and $\Delta T$ is the temperature gradient across the sample. For thin film structures, the temperature gradient is difficult to find out. We estimate the temperature gradient between heater and substrate using 3ω method and temperature gradient across the specimen is estimated using finite element modelling (FEM) (COMSOL). The temperature gradient between heater and far field temperature using 3ω method[42] is given by,

$$\Delta T = \frac{4 V_{3\omega}}{R' I_{rms}} \tag{7}$$

Where $V_{3\omega}$ is the third harmonic response, $R'$ is the resistance as a function of temperature and $I_{rms}$ is the heating current. The measured $R'$ is 0.07 Ω/K (Supplementary Figure S3). Using the 3ω method, we calculated the temperature gradient at heater to be 4.98 K, 10.9 K and 20.84 K for 15 mA, 20 mA and 30 mA of heating current respectively. Using FEM, we estimated the temperature gradient across the specimen to be ~14.08 mK corresponding to 20 mA of heating current. For modeling the temperature gradient, we assumed the $\kappa_{p-Si} = 25$ W/mK [43,44] and $\kappa_{Ni_{80}Fe_{20}} = 20$ W/mK [45]. For the temperature gradient, we calculated the $S_{LSSE}$ to be ~0.355 μV/K. This value is significant higher than the $S_{TSSE}$ reported for $Ni_{80}Fe_{20}$[10] thin film but lower than the $S_{LSSE}$ (0.8 μV/K[8]) reported for $Ni_{81}Fe_{19}$ thin films. It needs to be stresses that ISHE in this study is interfacial whereas the all the other reported studies use Pt for spin to charge conversion. From the S$_{LSSE}$ measurement this study, we can deduce that the $\theta_{SH}^{interfacial}$ is of the same order as $\theta_{SH}^{Pt}$. The calculated specimen temperature gradient is a function of the $\kappa_{p-Si}$. We repeated the calculation for $\kappa_{p-Si} = 20$ W/mK and 30 W/mK (Table S1-S2 and Supplementary Figure S4-S5) and the S$_{LSSE}$ is calculated to be 0.308 µV/K and 0.395 µV/K respectively. Phonons from sample and substrate are the primary component that governs the non-equilibrium state of metallic magnets (Py) where as magnons and phonons are responsible for non-equilibrium states in insulating magnets[46]. The SSE in semiconductors has been proposed to occur due to phonon drag but we observe a large SSE at 400 K. In order to ascertain the effect of phonons, we measured the second harmonic response as a function of temperature from 400 K to 10 K for an applied transverse in-plane magnetic field of 1500 Oe and 1T as shown in Figure 3 b for a second device. We do not observe effect of phonon drag and Si phonons in this measurement. We also measured the SSE at 200 K, 100 K and 20 K and the SSE is reducing gradually as the temperature is lowered as shown in Figure 3 c. From the temperature dependent study, we propose that the observed second harmonic response is attributed to the magnon mediated SSE. The second harmonic measurement shows a reduction as a function of temperature as expected for magnon mediated SSE, which further supports our assertion that observe behavior is SSE and not ANE.

From the experimental studies, we observe the SSE for transverse in-plane magnetic field and thermal SGE is observed for out of plane magnetization. The metal-semiconductor interface will lead to electron gas (EG) at the interface having thickness similar to the spin diffusion length as shown in Figure 3 d. The charge potential in the EG gives rise to strong Rashba SOC, which is the underlying cause of SSE, thermal SGE and SOT observed in this study. The spin orbit coupling due to lack of inversion symmetry in Si metal-oxide semiconductor field effect transistor (MOSFET) has been reported using magneto-transport behavior in two-dimensional electron gas (2DEG) [47-50] and using spin resonance measurements[51]. This agrees with the proposed Rashba effect behavior hypothesized in this study for the specimen having ferromagnetic metal-Si interface. The structure inversion asymmetric interface and intrinsic SOC are the essential requirements for strong Rashba SOC. In this study, large intrinsic SOC in $Ni_{80}Fe_{20}$[52] may give rise to the strong Rashba SOC as shown in Figure 3 d[53,54]. This hypothesis is further supported by observation of the strong Rashba spin split states at Bi/Si(111) interface[55]. In addition, the Rashba SOC mediated spin-Hall magnetoresistance has been reported in $Ni_{81}Fe_{19}$/MgO/p-Si thin films[56] and in n-Si[57]. The mechanistic explanation of the observed behavior is given in Figure 4. For the in-plane transverse magnetic field, the temperature gradient will generate a spin current leading to SSE at the interface as shown in Figure 4 a. While the out of plane magnetization will lead to spin accumulation causing a charge current across the interface due to thermal SGE. This charge current leads to SHE and resulting SOT observed in this study as shown in Figure 4 b[58]. We will like to stress that the thermal SGE behavior may originate due to currently unknown mechanism as well. Further experimental and simulation work is needed to develop mechanistic understanding of the behavior as well as the Rashba SOC responsible for it.

In conclusion, we observed giant spin-Seebeck effect and spin-orbit torques in $Ni_{80}Fe_{20}$/p-Si (poly) bilayer specimen. This measurement does not require any heavy metal for the spin to charge conversion. Instead, the inverse spin-Hall effect occurs at the $Ni_{80}Fe_{20}$/p-Si (poly) interface due to Rashba spin orbit coupling. The Rahsba spin-orbit coupling is proposed to occur due to electron gas at the interface. The electron gas behavior can be controlled using the Si semiconductor physics developed over decades. This may allow Si interfaces with giant spin-orbit coupling which may eclipse Pt as a primary spin detector. This may also lead to enhanced spin-Seebeck coefficient and in turn efficient thermal energy conversion. While the longitudinal spin-Seebeck coefficient measured in this study is similar to the values reported in literature but the room temperature $V_{SSE}$ observed in this study is one of the largest reported values [3,4,11,30] especially for a small temperature gradient of 10.9 mK across the interface. In addition to spin-Seebeck effect, the giant spin-orbit torque is also discovered, which is attributed to the thermal spin galvanic effect due to thermal spin pumping. The thermal spin-orbit torques lead to collapse of out of plane magnetic hysteresis of 25 nm thick $Ni_{80}Fe_{20}$ film. The thermal spin-orbit torques can be used to develop energy efficient memory devices utilizing the magnetization reversal behavior. In addition, these results will give impetus to the interfacial behavior at light elements having insignificant intrinsic spin-orbit coupling. These results bring the ubiquitous Si to forefront of spintronics research and will lay the foundation of energy efficient Si spintronics and Si spin caloritronics devices.

**Acknowledgement**

**Supplementary material**

The supplementary material includes the additional experimental measurements (specimen magnetoresistance, anisotropic magnetoresistance) and the details of COMSOL simulations carried out to estimate the temperature gradient across the specimen.

List of Figures:

Figure 1. a. The schematic of the experimental setup for the LSSE measurement, b. the false color SEM micrograph showing the device structure, c. the second harmonic response for applied magnetic field in transverse in-plane (y-direction) and out of plane (z-direction) directions and d. the second harmonic response as a function of heating power for an applied magnetic field of 1000 Oe (z-direction).

Figure 2. The second harmonic response related to the SSE as a function of magnetic field applied along y-direction and z-direction at 300 K for heating current of a. 15 mA, b. 20 mA, c. 30 mA and d. 50 mA. Arrows show the direction of magnetic field sweep.

Figure 3. a. The second harmonic response as a function of angular rotation of constant magnetic field of 2 T in the xy, zx and zy-planes and curve fitting showing a combined sine (SSE) and cosine (SOT) behavior, b. the second harmonic response as a function of temperature between 400 K – 10 K at applied magnetic field of 1000 Oe and 1 T, c. the second harmonic response as a function of magnetic field at 200 K, 100 K and 20 K, and d. the schematic showing the electron gas at the $Ni_{80}Fe_{20}$/p-Si (poly) interface ($\lambda_{SD}$ is spin diffusion length).

Figure 4 a. the schematic showing the mechanism of spin-Seebeck effect and b. the thermal spin-galvanic effect mediated thermal spin-orbit torque observed in this study.

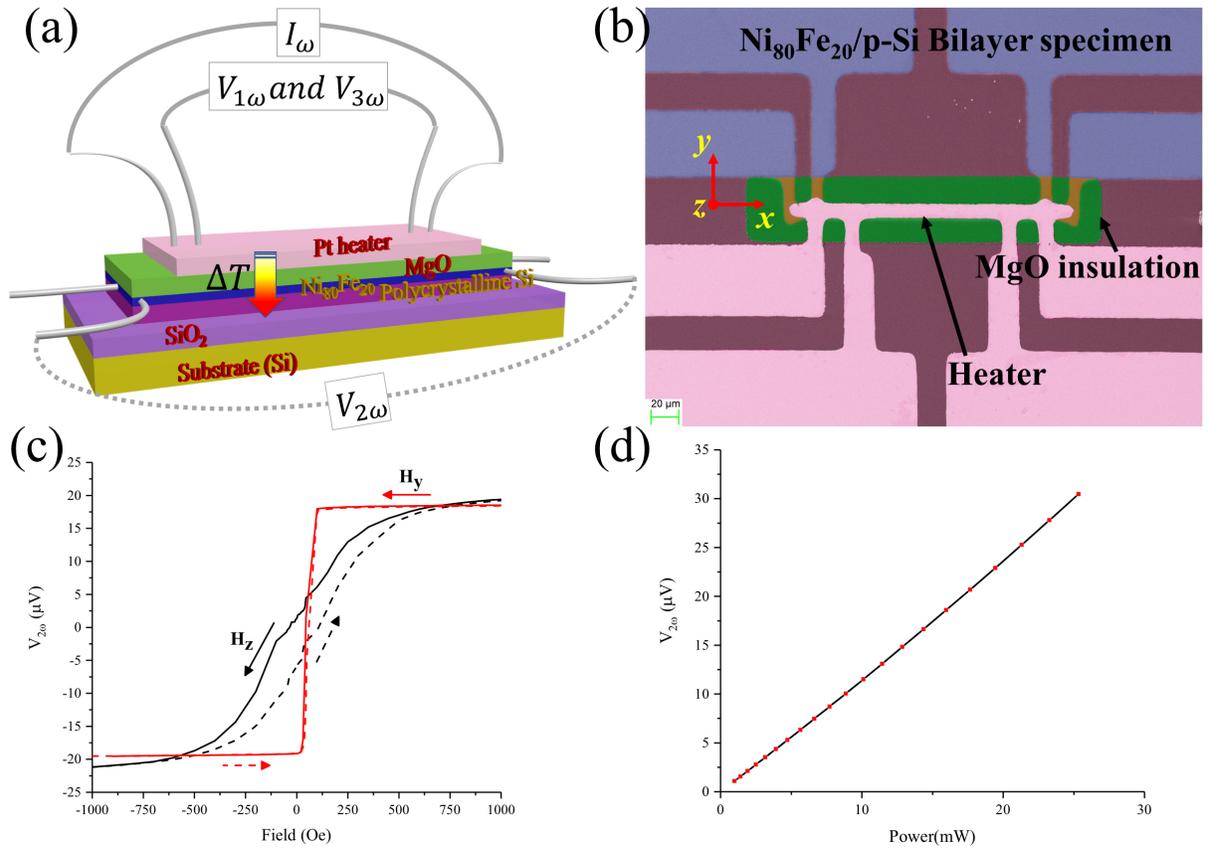

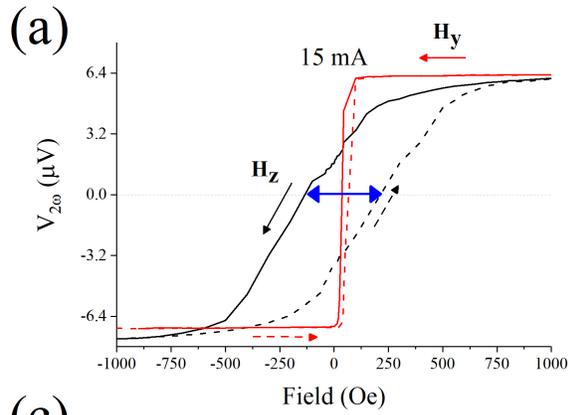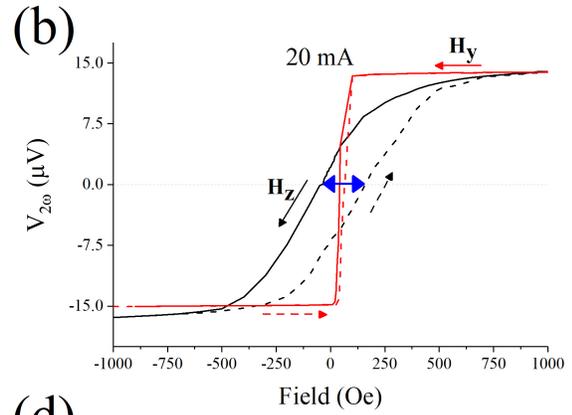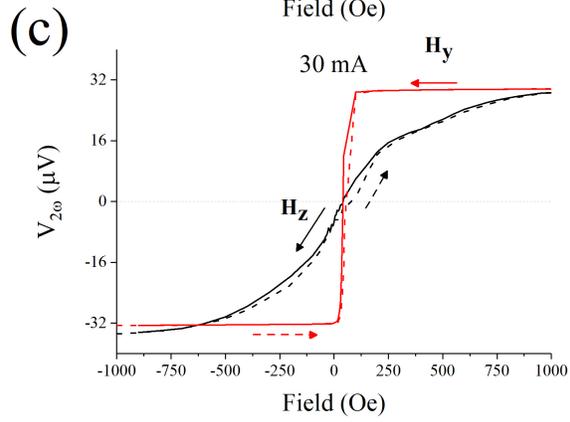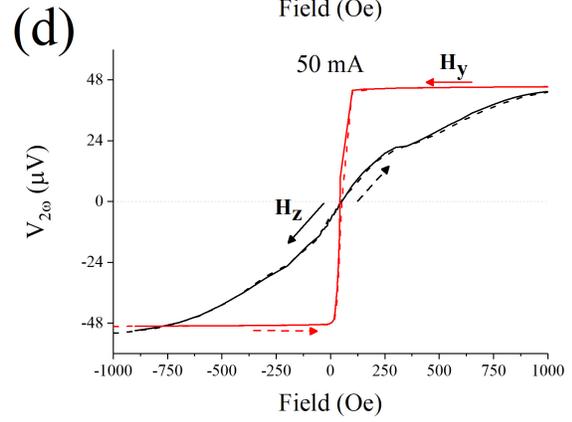

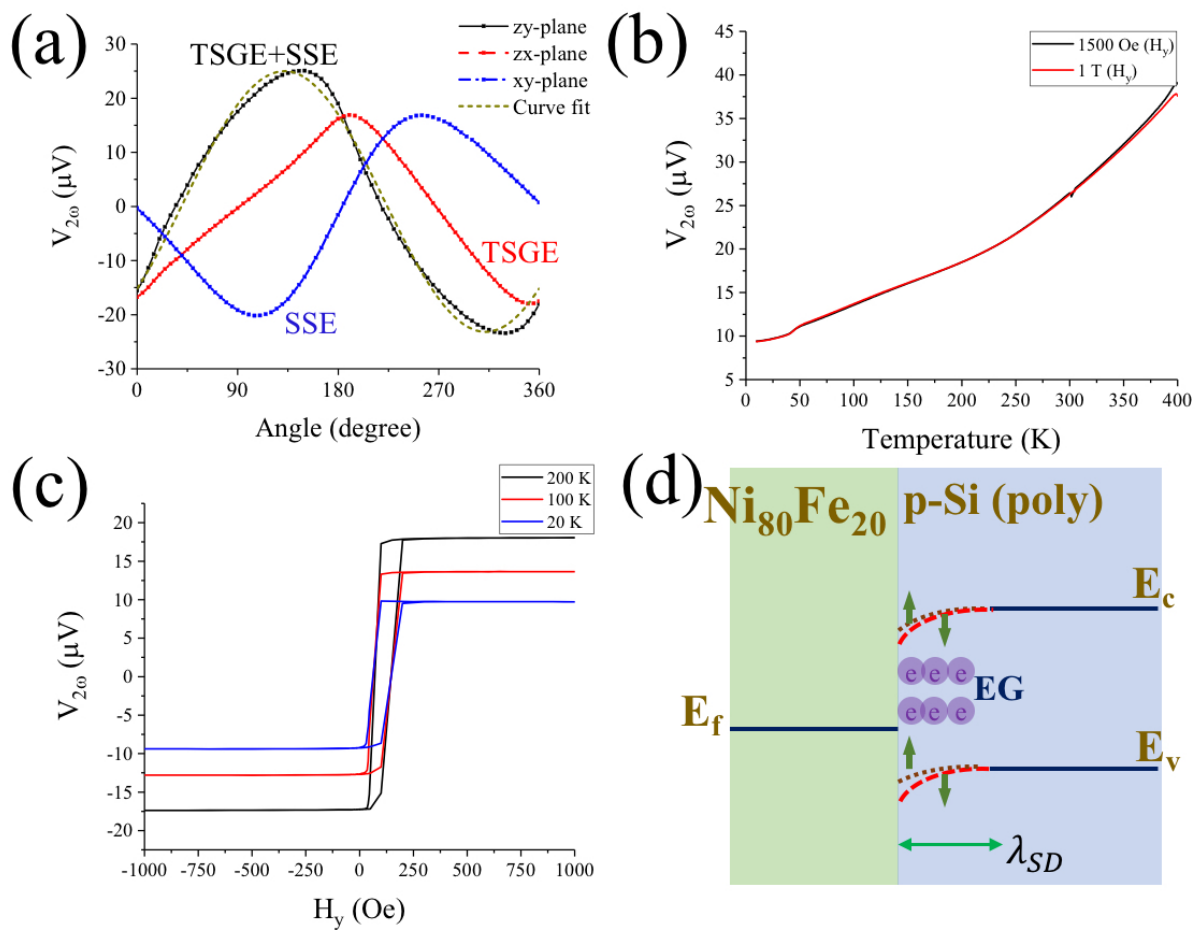

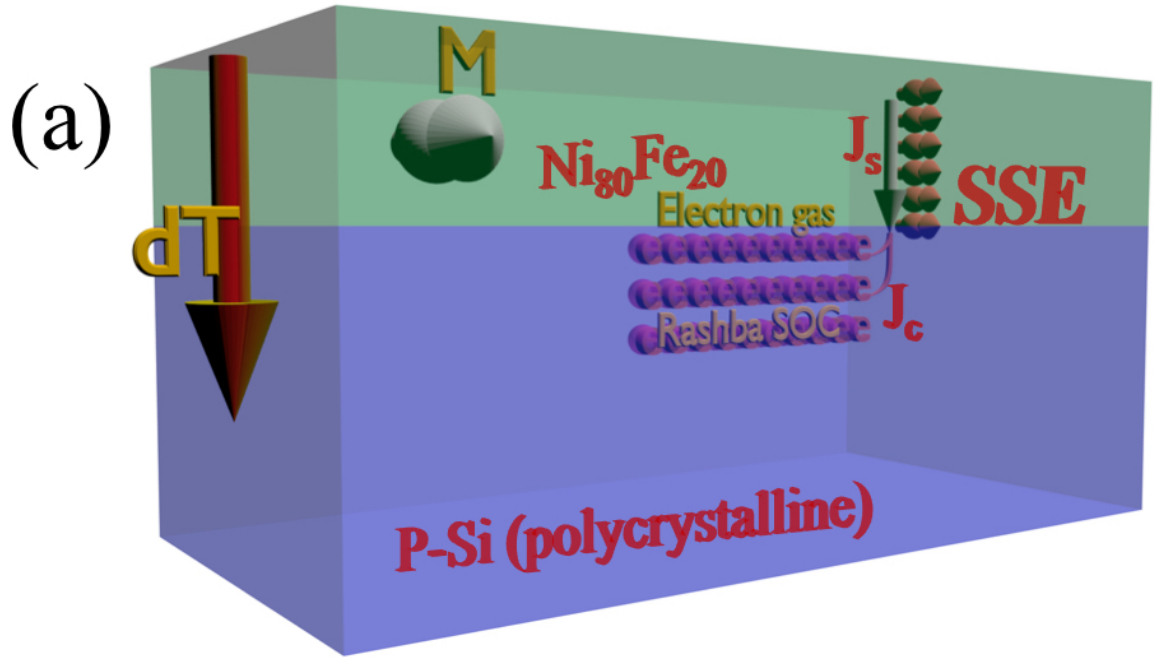

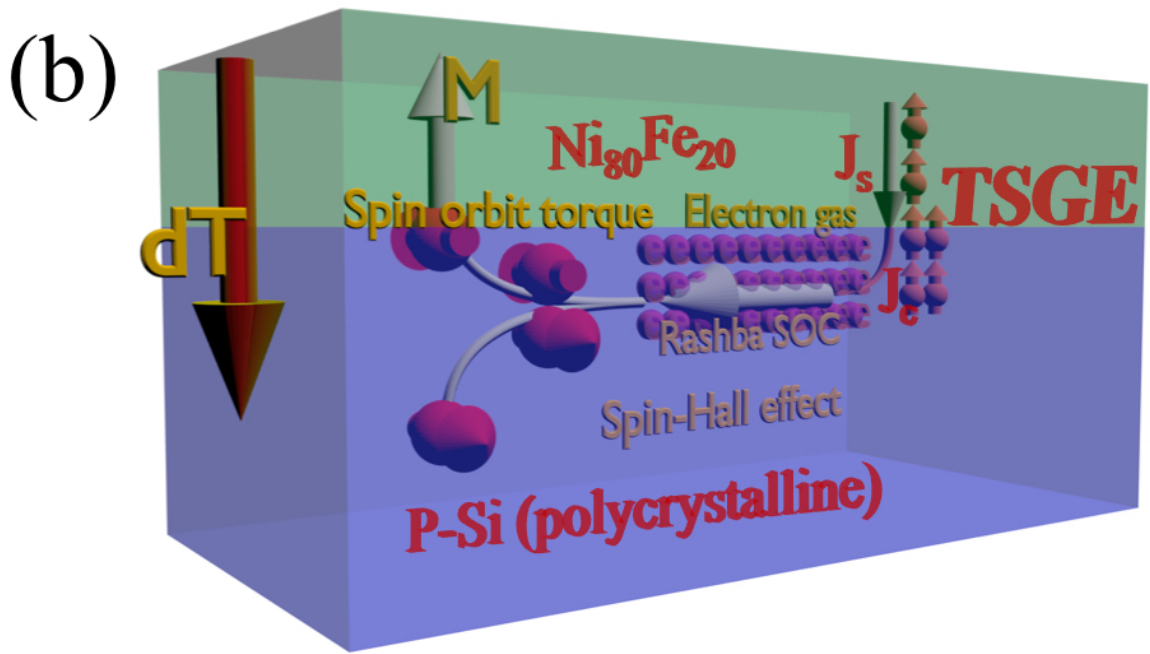

Supplementary Material:

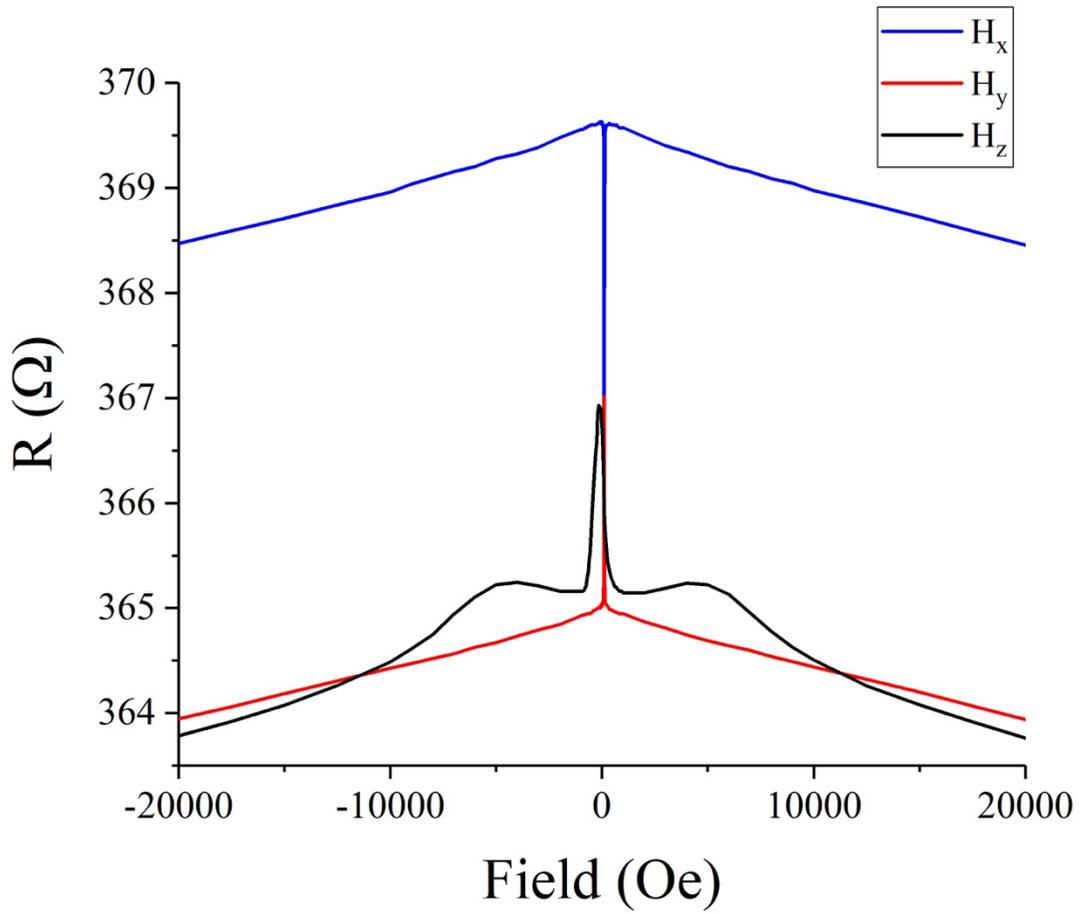

Supplementary Figure S1. The magnetoresistance measurement showing the easy and hard axis (z-direction) of the $Ni_{80}Fe_{20}$ thin film. The out-of plane (z-direction) saturation magnetization is approximately 1.25 T.

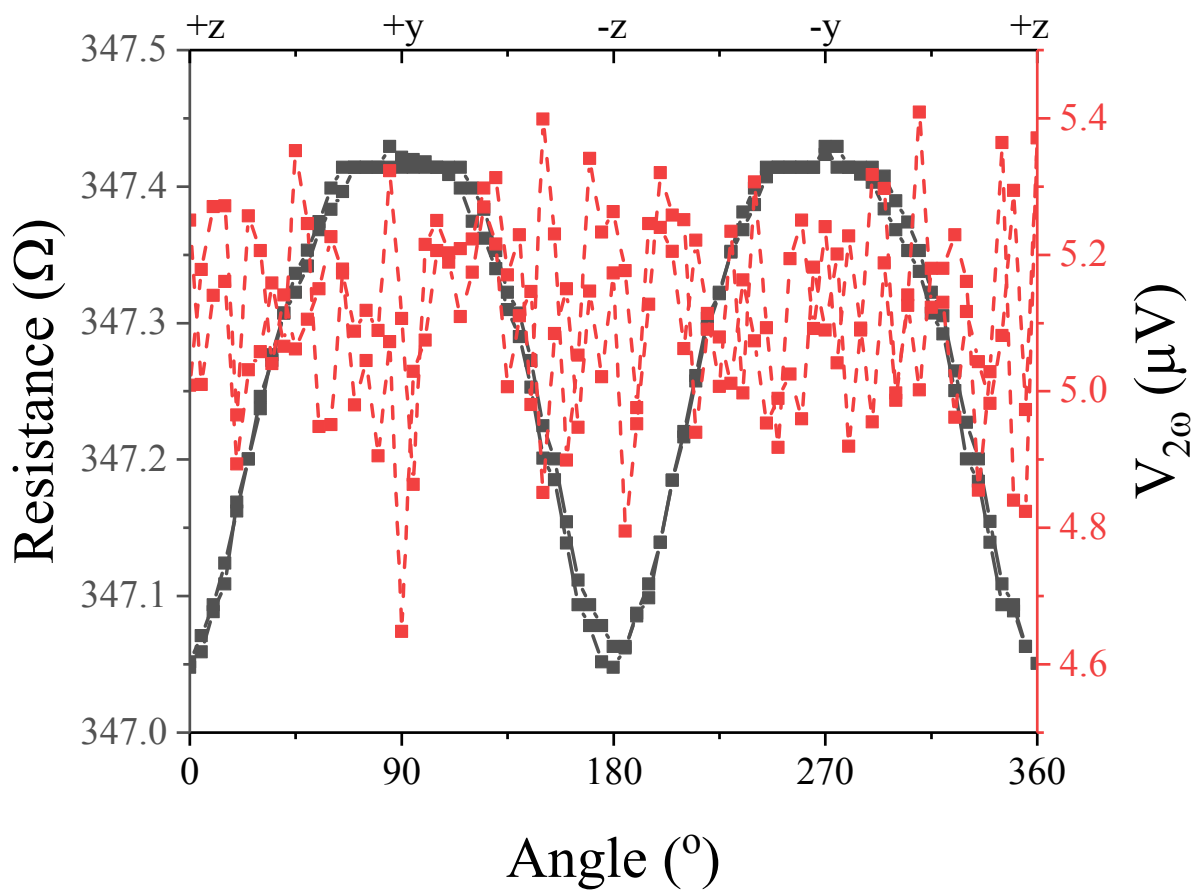

Supplementary Figure S2. The resistance and second harmonic response of the bilayer specimen for angular rotation of magnetic field in the yz-plane.

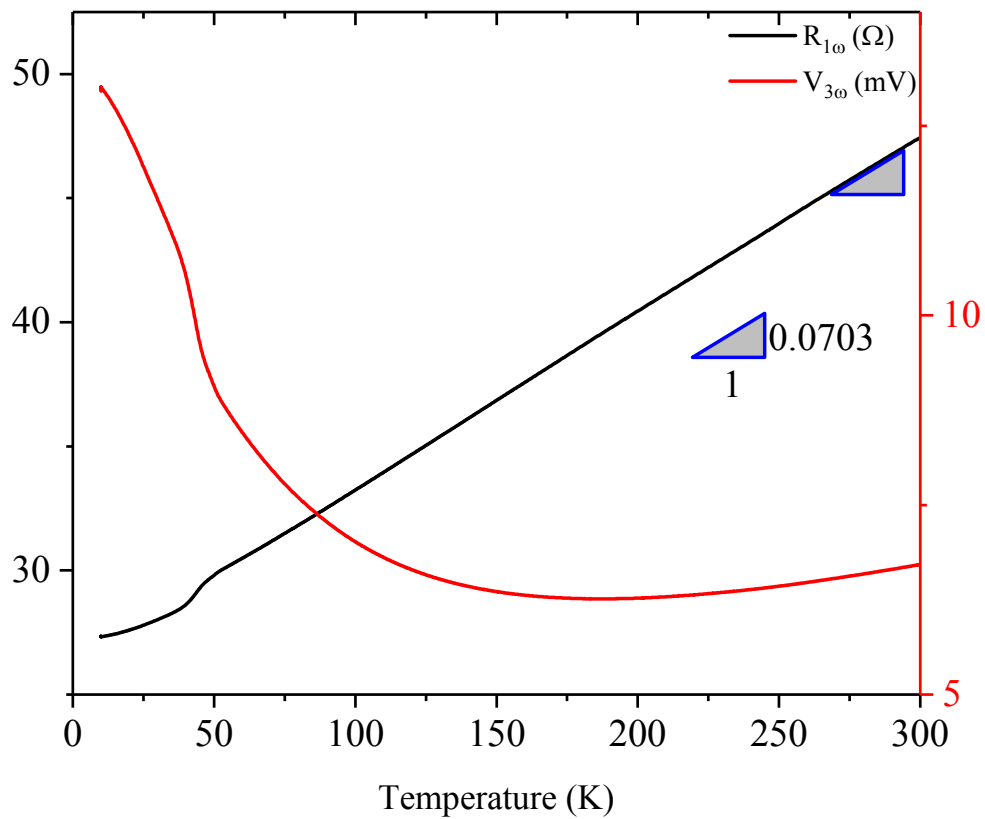

Supplementary Figure S3. The resistance and third harmonic response of Pt heater from 300 K to 10 K. The resistance data is used to fit and estimate $R' = \frac{dR}{dT}$.

Table S1. The properties used for the modelling

| Material | Thermal conductivity (W/mK) | Density (Kg/m$^3$) | Specific heat (J/KgK) |
|---|---|---|---|
| Platinum | 69.1 | 21450 | 130 |
| MgO | 30 | 3580 | 877 |
| Py | 19.6648 | 8740 | 502.415783 |
| p-Si | 22 | 2328 | 678 |
| Si | 130 | 2328 | 700 |
| SiO$_2$ | 1.3 – 1.5 | 2650 | 680 – 730 |

Table S2. The effect of $\kappa_{p-Si}$ on the temperature gradient across the specimen

| Thermal conductivity (W/mK) of p-Si | Heater Temperature (T1) | Temperature difference |
|---|---|---|
| 20 | 304.98 K | 0.00741716 K |
| | 310.9 K | 0.01623436 K |
| | 320.84 K | 0.03103890 K |
| 25 | 304.98 K | 0.00643506 K |
| | 310.9 K | 0.01408479 K |
| | 320.84 K | 0.02692908 K |
| 30 | 304.98 K | 0.00578012 K |
| | 310.9 K | 0.01265127 K |
| | 320.84 K | 0.02418830 K |

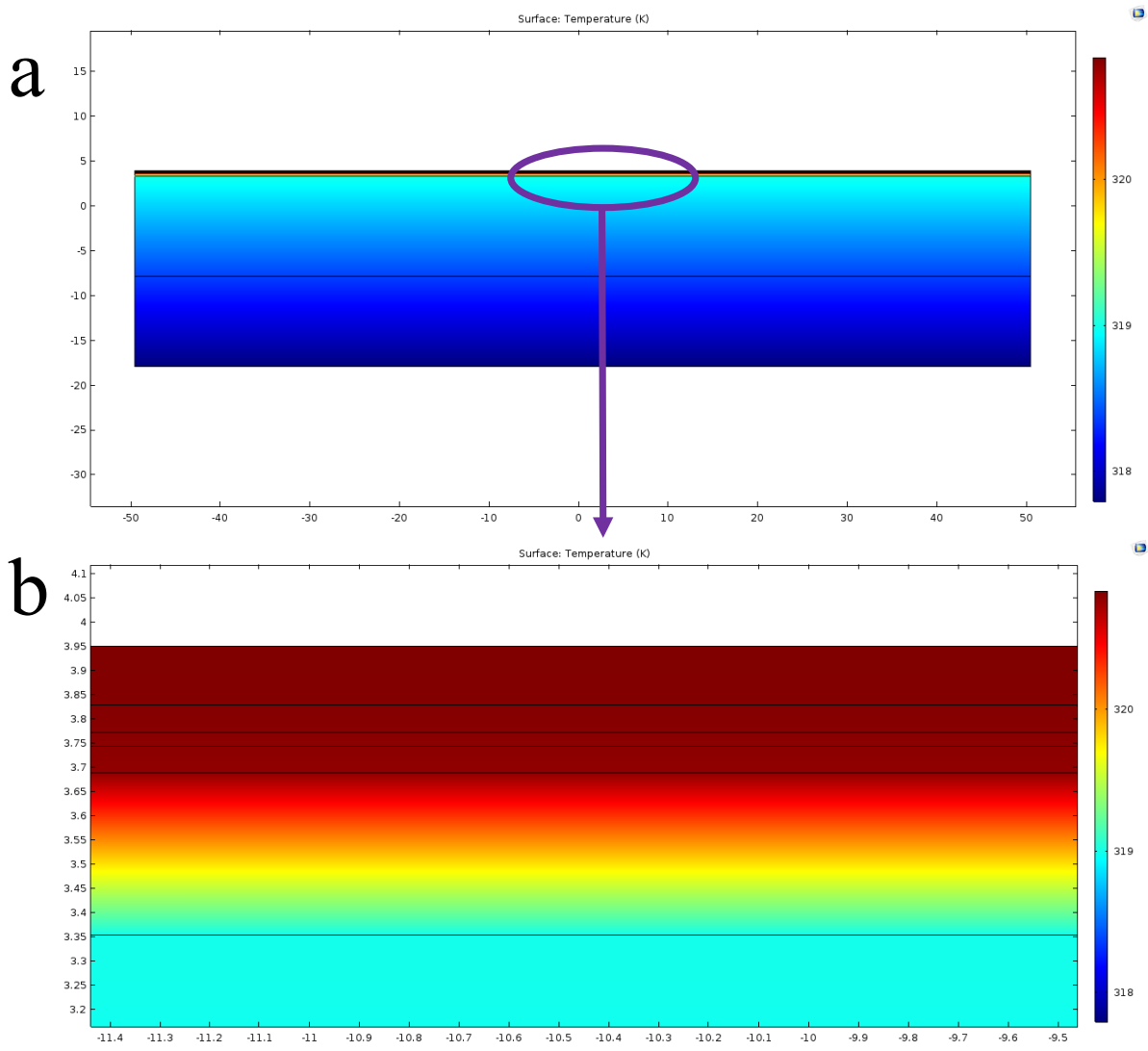

Supplementary Figure S4. A. the COMSOL model showing the temperature gradient between the heater and the substrate and b. the temperature gradient across the layered structure for 30 mA of heating current.

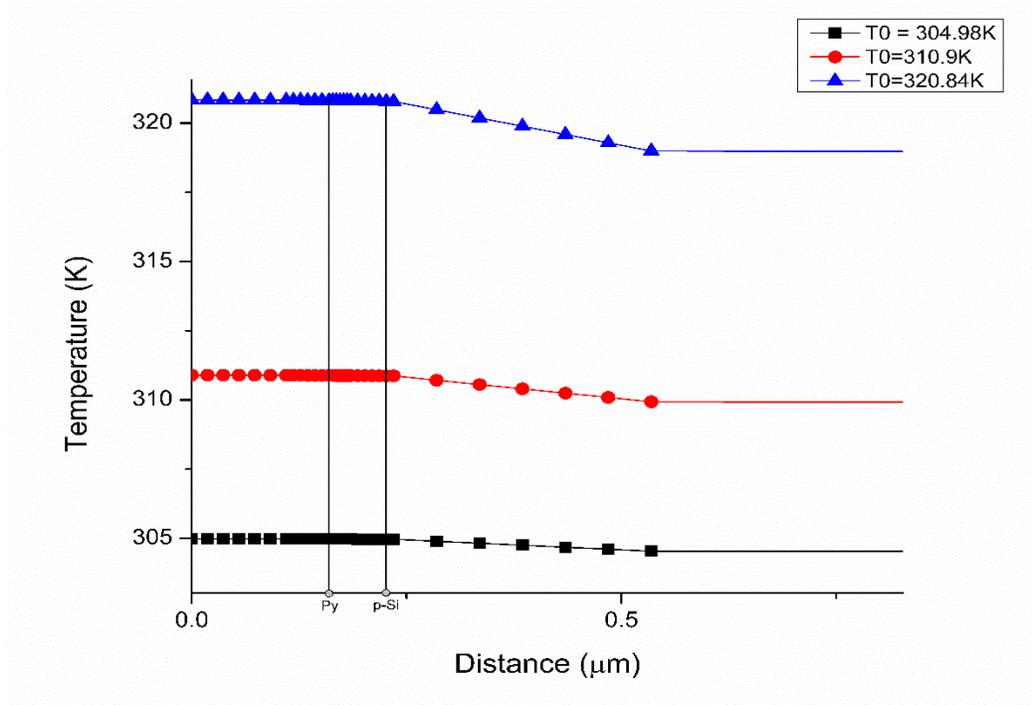

Supplementary Figure S5. The temperature gradient across the specimen for heater temperatures corresponding to 15 mA, 20 mA and 30 mA of heating current.